\documentclass[runningheads]{llncs}
\usepackage{amsmath,amsfonts}
\usepackage{algorithmic}
\usepackage{algorithm}
\usepackage{graphicx}
\usepackage{textcomp}
\usepackage{xcolor}
\usepackage{subfigure}
\usepackage{comment}
\usepackage{multirow}
\usepackage{pdflscape}
\usepackage{tabularray}
\usepackage{framed}
\usepackage{hhline}
\usepackage{bbding} %
\usepackage[most]{tcolorbox}
\usepackage{tikz}
\usepackage{filecontents}
\usepackage{hyperref}
\usepackage[title]{appendix}
\usepackage{bbding}
\begin{document}
\title{E-DoH: Elegantly Detecting the Depths of Open DoH Service on the Internet}
\author{Cong Dong\inst{1} \and
Jiahai Yang\inst{1,2} \and
Yun Li\inst{3,4} \and
Yue Wu\inst{1,2} \and
Yufan Chen\inst{3,4,(}\Envelope\inst{)} \and
Chenglong Li\inst{1,2,(}\Envelope\inst{)} \and
Haoran Jiao\inst{1} \and
Xia Yin\inst{1,2} \and
Yuling Liu\inst{3,4}
}
\authorrunning{Cong Dong et al.}
\institute{Zhongguancun Laboratory \\
\email{dongcong,lichenglong,jiaohaoran@zgclab.edu.cn}\\
\and
Department of Computer Science and Technology, Tsinghua University\\
\email{yang@cernet.edu.cn,wuyue23,yxia@tsinghua.edu.cn} \\
\and 
Institute of Information Engineering, Chinese Academy of Sciences\\
\and
School of Cyber Security, University of Chinese Academy of Sciences
\email{liyun1996,chenyufan,liuyuling@iie.ac.cn}
}
\maketitle              %
\begin{abstract}
In recent years, DNS over Encrypted (DoE) methods have been regarded as a novel trend within the realm of the DNS ecosystem.
In these DoE methods, DNS over HTTPS (DoH) provides encryption to protect data confidentiality while providing better obfuscation to avoid censorship by multiplexing port 443 with web services.
This development introduced certain inconveniences in discovering publicly available DoH services.
In this paper, we propose the E-DoH method for elegant and efficient DoH service detection. 
First, we optimized the probing mechanism to enable a single DoH connection to accomplish multiple tasks including service discovery, correctness validation and dependency construction. Second, we propose an efficient DoH detection tool. This tool can enhance probing efficiency while significantly reduce the required traffic volume. Third, based on the above optimization methods, we conducted an exploration of the IPv4 space and performed an in-depth analysis of DoH based on the collected information.
Through experiments, our approach demonstrates a remarkable 80\% improvement in time efficiency, and only requires 4\%-20\% traffic volume to complete the detection task.
In wild detection, our approach discovered 46k DoH services, which nearly doubles the number discovered by the state-of-the-art. Based on the collected data, we present several intriguing conclusions about the current DoH service ecosystem.
\end{abstract}
\section{Introduction}

Measuring DoE services has captured the attention of current researchers. The primary objective of these measurement endeavors is to discover DoE services in the wild. Among the various DoE measurement studies, the measurement for DoH services gathers more significant attention than others. In addition to the encryption protection gained in data transmission, DoH services can further enhance DNS service obfuscation by multiplexing with the traditional 443 port. This feature increases the difficulty of discovering DoH services, as the mere presence of an open port does not conclusively indicate the provision of DoH services. Therefore, the measurement of DoH services become a focal point in current DoE measurement research.

\textbf{Previous Shortage} Previous research endeavors have introduced approaches for detecting open DoH services in the wild without relying on published lists. Their experimental findings have demonstrated the feasibility of discovering wild DoH services. Nevertheless, there are two evident shortcomings of them. 
First, previous approaches only probed the availability of services. Their analysis of DoH is limited to the TLS and HTTP layers, lacking an assessment of the quality of service at the DNS level and a comprehensive evaluation of DoH in the overall DNS ecosystem. Second, the previous measurement methods require a significant amount of time. In order to discover DoH services as comprehensively as possible, previous research has adopted the enumeration strategy, which extensively enumerates the TLS fields as well as HTTP fields that provide DoH services to achieve a successful handshake for the underlying protocol. However, this enumeration strategy incurs significant time costs. Additionally, their enumeration method also results in a large amount of traffic volume. In practical applications where a large number of 443 ports are hosting web services, the issues of probing efficiency and traffic volume scale become more unacceptable.

\textbf{Our Work}
In this paper, we bridge the aforementioned research gaps by proposing a novel elegant DoH (E-DoH) measurement method. The method can efficiently realize the in-depth measurement of the DoH ecosystem. In summary, we tackled two challenges to bring this method. First, drawing inspiration from the in-depth measurement of the traditional DNS ecosystem, existing methods require at least two probing processes to assess service availability, correctness, and dependencies. At the same time, complex manual configurations should be conducted on the controllable authoritative server for the measurement. To address this challenge, we introduce a probing mechanism based on wildcard domain names. By mapping unique probing domain names for each target to a single wildcard domain name in the zonefile, this mechanism requires only one DoH query to accomplish an in-depth exploration of the DNS service ecosystem, with minimal manual configuration on the backend. Second, detecting DoH requires successful handshakes at both the SSL and HTTP protocol layers. Previous methods used a brute-force enumerative strategy and script languages, resulting in suboptimal probing efficiency and a large amount of traffic volume. To address this challenge, we propose the dynamic protocol negotiation strategy. The dynamic negotiation handshake method reuses successfully negotiated lower layer protocols and dynamically inferring higher layer parameters. Thus it can improve detection efficiency while reducing unnecessary traffic volume. In the specific implementation, this strategy is integrated into Golang, leveraging the advantages of compiled languages in multithreading execution to enhance detection efficiency. In experiments, we evaluate the detection performance of the tool based on three datasets. Results show that our tool can achieve an 80\% improvement in time efficiency, and reduce unnecessary packet transmissions by 80\%-96\%. 

\textbf{Key Findings} Based on the proposed E-DoH method, we performed measurements of the DoH ecosystem in the IPv4 space. In total, we discovered 46k DoH services, which nearly doubles the number discovered by the state-of-the-art \cite{li2023longitudinal}. As an in-depth investigation of the ecosystem, there are three observations.
First, the majority of current DoH services are not open to the public, with only about 5k DoH services returning valid responses. These DoH services exhibit high service quality, with 98\% of these 5k DoH services returning correct answers. 
Second, 92\% of the DoH services are forwarding resolvers rather than recursive resolvers that directly perform iterative query operations. Compared to traditional resolvers, only 12.5\% are forwarder recursive \cite{parkLargeScaleBehavioralAnalysis2022}. This trend indicates that DoH resolvers and traditional resolvers have distinct roles within the DNS ecosystem.
Third, there are complex dependency relationships between DoH forwarding resolvers and the backend recursive resolvers, showing high levels of clustering for both forwarding and recursive resolvers. This indicates that in the DoH ecosystem, multiple components are interconnected.

\section{Preliminaries}

\begin{table}
\centering
\label{Tab_compare}
\caption{Comparison of previous studies on resolver detection}
\resizebox{\linewidth}{!}{
\begin{tblr}{
  cells = {c},
  vline{2} = {-}{},
  vline{2} = {2}{-}{},
  hline{1-2} = {-}{},
  hline{2} = {2}{-}{},
  hline{8} = {-}{0.08em},
}
                                            & Highlights                                              & DoH Support & Number   & Year & Shortage                                      \\
\cite{parkLargeScaleBehavioralAnalysis2022} & {Subdomain encoding, \\backend Log}                      & \XSolidBrush    & -                      & 2022 & {Low efficiency, \\Multiple requests required} \\
\cite{izhikevichZDNSFastDNS2022}            & {High detection efficiency, \\DNS packet customization} & \XSolidBrush    & -                      & 2022 & Not support for DoH                           \\
\cite{luEndtoEndLargeScaleMeasurement2019}  & {Passive Traffic Analysis, \\First measurement for DoH} & \Checkmark      & 61                     & 2019 & Incomplete coverage                           \\
\cite{luoMeasurementEncryptedOpen2022}      & Host name collection                                  & \Checkmark      & 5715                   & 2022 & Low efficiency, Manual effort                               \\
\cite{garciaLargeScaleAnalysis2022}         & 6 DoH requests (2 versions,3 methods)                                          & \Checkmark      & 4354                 & 2022 & Low efficiency, Large traffic volume, Empty Query                   \\
\cite{li2023longitudinal}                   & 24 DoH requests (2 versions, 3 methods, 4 paths)                                       & \Checkmark      & 25970 & 2023 & Low efficiency, Large traffic volume, Empty Query                   
\end{tblr}
}
\end{table}

\subsection{Encrypted resolver measurement}
In the realm of (DoE) measurements, current research predominantly focuses on service discovery \cite{luEndtoEndLargeScaleMeasurement2019,luoMeasurementEncryptedOpen2022,garciaLargeScaleAnalysis2022,jerabekAnalysisWellKnownDNS2023a}. \cite{garciaLargeScaleAnalysis2022} introduces a robust three-phase DoH detection method by active probing. As a result, \cite{garciaLargeScaleAnalysis2022} identifies 4354 DoH services. 
Building upon this foundation, \cite{luoMeasurementEncryptedOpen2022} enhances the discovery by providing domain names in the URL field of the HTTP protocol. Their domain names of the IP addresses are from the certificate SNI or PTR record of the IP. As a result, this domain name enhancement method can discover 5715 DoH services. Noteworthy is the observation that despite incorporating domain names, 87.4\% of machines are found to access DoH servers solely via IP addresses, while the remaining 12.6\% utilized both IP addresses and domain names. In addition to exploring the domain name, \cite{li2023longitudinal} introduces a method for enumerating the path field in the HTTP filed. As a result, \cite{li2023longitudinal} discovers approximately 26k DoH services.
A common challenge among the aforementioned active detection methodologies is the requirement for multiple complete network connections to accurately identify potential DoH services. This results in decreased efficiency and places additional strain on operational networks.
To address the issues, our paper proposes a novel dynamic protocol negotiation approach. By reusing established connections and dynamically inferring upper-layer protocols, our method enhances efficiency and reduces traffic volume transmission overhead. This innovative approach aims to alleviate the efficiency concerns associated with repeated complete network connections while minimizing the impact on operational networks.

\subsection{Traditional resolver measurement}
In addition to service discovery, traditional resolver measurements also focus on assessing the status of resolve services \cite{parkWhereAreYou2019,parkLargeScaleBehavioralAnalysis2022,nawrockiTransparentForwardersUnnoticed2021a,izhikevichZDNSFastDNS2022,changZMapPerformanceOpen2022a,10152616}.
Regarding measurement tools, ZDNS \cite{izhikevichZDNSFastDNS2022} is a powerful open source tool for traditional DNS measurements to support the data collection work of the OpenINTEL project. However, this tool only provides measurements for the traditional DNS service.

To achieve an in-depth exploration of the DNS service ecosystem, deploying a self-controlled domain is necessary in the measurement process. \cite{parkWhereAreYou2019,parkLargeScaleBehavioralAnalysis2022} propose assigning and requesting unique subdomains for each probing target to achieve correctness and dependency measurements simultaneously. However, this method requires a significant amount of configuration in the zonefile.
Besides, loading this massive zonefile into memory often leads to memory crashes. In order to address the issues arising from the aforementioned approach, \cite{10152616} proposed to optimize this process by separating correctness and dependency measurements into two distinct steps. Thus it eliminate the need to configure correctness records for each domain in the zonefile. However, the accompanying cost is to make two requests for each target.
The probing mechanism introduced in this paper utilizes wildcard domain names, allowing the simultaneous measurement of correctness and dependency of open resolvers without the need for configuring extensive zonefile. Moreover, this method is not limited to DoH. It is equally applicable to the detection of other types of resolvers, such as traditional DNS services, DoT and DoQ.

\section{Overview of Measurement}

The measurement of DoH service can be divided into four steps. 
The first step is to scan open port 443 to identify potential DoH service providers. 
The second step involves deploying the registered public domain names onto a controlled server machine and configuring the associated wildcard subdomains by managing zonefile. 
The third step entails sending DoH queries to the hosts identified in the first step and collecting the returned responses by our proposed tool. 
The fourth step involves collecting probing returned results and nameserver request logs for further analysis.

During the measurement process, the main challenges lie in \emph{1) how to accomplish multiple DoH service status measurement tasks in a single DoH request} and \emph{2) how to efficiently perform DoH request probing for a large number of targets}. To address these issues, this paper proposes to optimize previous detection methods from measurement mechanism and probing implementation.
In the measurement mechanism, the deployment of controlled domains and the wildcard domain mechanism play a crucial role. Each detection target is assigned with a unique domain name, while this name is also mapped to the wildcard domain record configured in the zonefile. Therefore, it becomes possible to collect multiple information about DoH service status through a single DoH request. 
In probing implementation, optimization is achieved in two aspects. First, at the strategy level, the dynamic protocol negotiation strategy is proposed for fined grained protocol control and reusing. Second, at the implementation level, the probing program is efficiently implemented using the concurrent language Golang. Additionally, error timeouts are implemented based on thread execution time, reducing the occurrence of long-running zombie threads caused by network connection errors.

\section{Detection Mechanism}

\subsection{Controlled Server Deployment}
The measurement procedure requires deploying a controlled authoritative server for authenticity verifying and query logging. This trick is used by several recent DNS measurement works. Most DNS measurement works build their own authoritative server rather than requesting a public domain, like example.com. 
We follow the routine that building our own authoritative server based on three considerations. First, the controlled domain name facilitates the management of subdomains. It is convenient to append new subdomains by modifying the zonefile. Second, the controlled domain name ensures the correctness verification. Since the value of the controlled domain and its subdomain are managed by ourselves, the authenticated value can be verified. Third, the controlled domain name can avoid the potential service pressure to public domains. In wild exploration, if there are a significant number of active DoH services, probing them using publicly registered domain names could exert substantial load pressure on those domain names, resembling a DoS (Denial of Service) attack.
Thus, we register a domain name \emph{uniquetest.today} under the \emph{.top} TLD with the  \emph{de facto standard} bind9 as the resolver backend software.

\subsection{Request Domain \& Backend Logs}

\begin{figure}
\centerline{\includegraphics[width =0.6\linewidth]{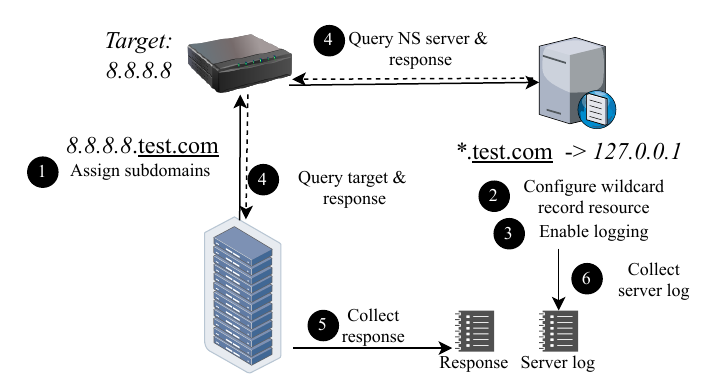}}
\caption{Domain assigning and backend logging}
\label{Fig_assign}
\end{figure}
For each target, the tool sends a unique domain in the DoH packet and expects a correct response from the target. Instead of requesting a random domain name, we organize the requested domain names to maximize information retrieval in a single DoH request. In short, this can be achieved by following steps, 1) assigning unique subdomains to each detection target for probing, 2) mapping these subdomains to a wildcard resource record, 3) enabling the logging function of the backend framework, 4) sending DoH requests to probing targets, 5) collecting responses from the prober and logging records from the backend framework.

The detailed detection procedure is as follows. 
First, each target is assigned a unique subdomain under the applied domain. The assigned subdomain is filled into the DoH message during the subsequent detection process to distinguish between different targets. Each assigned domain name consists of two parts, the first part is the target IP, and the second part is the applied domain. To avoid multiple iterative requests caused by '.' of the target IP in the first part, we replace the '.' with '-' as the final first part.
Second, all allocated subdomains are mapped to a single wildcard resource record in the authoritative server zonefile. As opposed to a corresponding resource record for each allocated subdomain, the wildcard mechanism reduces the cumbersome tasks of creating corresponding records for each detection subdomain and alleviates the substantial memory load when loading corresponding zonefile. As for the configured value of the wildcard record, a single non-public address is chosen based on identical consideration for further response correctness verification.
Third, the logging function of the backend recording framework is enabled. After enabled, the framework will record the IP address and the domain name from which the request originated when the domain request arrives at the authoritative server. 
Fourth, the detection DoH message is sent to targets with the customized probing implementation. Specific details of the implementation are presented in V Probing Implementation. 

To provide a more intuitive explanation of the procedure, an example is provided as shown in \autoref{Fig_assign}. If a target with IP 8.8.8.8 is recognized as providing service on 443 port, this target is assigned with an IP 8-8-8-8.uniquetest.today. At the same time, a wildcard resource record is appended to the zonefile. With this record, any subdomain names under unqiuetest.today can be mapped to 127.0.0.1. Moreover, the real resolver that connects to the authoritative server will be logged by the backend framework with the requested subdomain name. Finally, the responses received by the prober are correlated with the backend logging records for further analysis.

\section{Probing Implementation}

\subsection{Dynamic Protocol Negotiation}

In traditional DNS service detection, the preliminary assessment of which hosts offer DNS services can be determined by the open 53 port. However, this cannot be directly adapted for DoH service detection. Since the DoH service is indistinguishable from a general web service in that both use port 443 to provide web services. 
Therefore, the detection of DoH requires a successful DoH session with the 443 port of targets to conclusively confirm it as a DoH service rather than a web service.

Although RFC 8484 provides standard specifications that DoH services should adhere to at each layer, practice users may violate these specifications due to non-standard configurations and privacy considerations. We categorize these specifications into two parts. One part of the specifications can be determined through negotiation parameters at the lower protocol level, such as determining the \emph{HTTP version} through the TLS ALPN field. The other part of the specifications requires confirmation of the correct parameters through enumeration, such as the \emph{path} in the HTTP header. By specifying the two specification, we propose the fine grained protocol negotiation and reuse of established underlying connections to reduce traffic volume.

Previous approaches try to complete a successful DoH communication by enumerating all possible specifications by initiating different complete connections. In probing, previous approaches require initiating several separate network connections to transmit probing DoH packets, requiring a complete process from TCP to HTTP connection for each DoH packet transfer. In contrast, dynamic protocol negotiation can detect the service in a single network connection. First, dynamic protocol negotiation can specify the HTTP version from the TLS ALPN field. In subsequent detection of HTTP methods and path fields, dynamic protocol negotiation can reuse the previously established TLS connection to transmit probing HTTP packets. Thus it requires only one process from TCP handshake to HTTP connection establishment for all DoH packet transfer.

\subsubsection{TLS Negotiation}

In TLS negotiation, we configure to accept all TLS versions from TLS 1.0 to TLS 1.3. Although \cite{hoffmanDNSQueriesHTTPS2018} specified that DoH should use TLS 1.3 for connection, it is found that practice DoH services are still transmitted via TLS 1.2 or even TLS 1.0 \cite{luoMeasurementEncryptedOpen2022,li2023longitudinal}.
We hypothesize that the reason for using a lower version of TLS is due to different DoH deployment methods. In general, there are two main deployment ways for DoH services. The first is to deploy the DNS service software, like bind9 directly on the 443 port. Such a deployment exists only in DNS service software in the backend. Thus underlying services including TLS, HTTP, and DNS are all handled by the deployment DNS service software. The second type is to deploy the DNS software to the backend of the application server, such as Node.js. This deployment way is used to share the 443 port with other services. For example, if some services need to provide more than one service on the 443 port, the services will be forwarded to the HTTP layer by the server of Node.js according to the URL or path field. In order to be compatible with the previous services, a lower version of the TLS protocol that was previously deployed is thus reused. In other words, TLS services and HTTP services can be provided by the application server, while DNS services are provided by the DNS service software in this deployment. This deployment approach therefore results in DoH services being provided over a lower version of the TLS protocol. 

\subsubsection{HTTP Negotiation}
HTTP offers a variety of control fields as specifications. It is necessary to systematically enumerate and explore these fields. First, the HTTP version of a subset of targets can be determined through fined protocol control.
\emph{Application Layer Protocol Negotiation (ALPN)} is an extension field of TLS, it can be used to select the next stage of HTTP requests. By providing both h2 and http1.1 in TLS handshake, most targets will respond to the probing clients with specific selection. However, a few of them will not respond to this field. In such a situation, both h2 and http1.1 are tried sequentially on the established TLS.
Second, the HTTP request method field is determined through enumeration. GET/POST/JSON methods are sequentially sent to the target. While RFC suggests that DoH services should provide both GET and POST methods, we also include the commonly used JSON method in web services for comprehensiveness.
Third, the HTTP request URI field is similarly determined through enumeration. Although RFC recommends using "/dns-query" as the standard DoH URI, this field is employed by service providers as one of the authentication conditions in practice. By setting unusual URIs, server providers can restrict the scope of service provision. Through the frequent pattern mining of URI templates in publicly available DoH lists, we used four URI templates for detection, which are "/dns-query," "/resolver," "/doh," and "/."

\subsection{Implementation of Golang}

From the perspective of task characteristics, the service detection process is I/O-intensive rather than CPU-intensive. Besides, due to network latency, the I/O in the detection process is more inefficient. Most of the time in the detection process is spent waiting for network responses from the target. Hence, minimizing I/O waiting time is crucial for optimizing the overall implementation. In summary, we introduce the parallel detection mechanism within Golang to reduce the time for individual targets.

The parallel implementation is carried out using the Golang language. The overall probing efficiency can be improved by allocating the probing tasks to multiple threads through the parallel mechanism to reduce the waiting time. Inspired by the architecture of ZDNS, we also use the Golang language for overall implementation.
In particular, when determining the exact number of threads, it is not necessarily the more threads the better. A large number of threads will pose the problem that execution time is insufficient for each thread, resulting in errors, such as packet loss or parsing errors. Therefore, the exact number of threads should be determined by taking into account various factors such as machine performance and bandwidth. In this paper, we determines the optimal number of threads by conducting experiments on a list of publicly available DoH services as shown in \emph{6.2 Thread Tunning}.

\section{Efficiency Experiment}

\subsection{Single Thread Comparison}

In order to show the performance of this method in terms of detection efficiency, three public DoH lists are used for evaluation. Two of the datasets are from \cite{garciaLargeScaleAnalysis2022}, known2021 and known2022 respectively. However, due to the passage of time, there is a risk of IP address changes or service disruptions in these two datasets. Therefore, in this paper, we have crawled from the list of publicly available DoH services to produce the \emph{known2023} dataset \cite{bartDNSoverHTTPSDOHProvider2023,DNSCryptListPublic,PublicResolversDnsprivacy}. Results are shown in \autoref{Tab_efficient_results}.

\textbf{Compared Methods}
Totally, we introduced 7 methods for comparison. 
The first method is dig-9.18.12, which is an adjunct to the well-known DNS software, denoted as bind9 (DIG).
The second method is a python-based DoH detection with only the port 53 opening \cite{deccioDNSPrivacyPractice2019} denoted as D-C53. 
Methods 3-5 are proposed in \cite{garciaLargeScaleAnalysis2022}. We split their two-phase DoH detection into three independent methods, denoted as G-NSE, G-VAL, and G-ALL. 
The sixth method is a python-based DoH detection with domain name as enhancement from \cite{luoMeasurementEncryptedOpen2022}, denoted as L-DE. 
The seventh method is a python-based DoH detection with path enumerating from \cite{li2023longitudinal}, denoted as L-LC.

\textbf{Efficient Comparison} First, E-DoH offers significant performance advantages. Roughly, E-DoH only requires 5\% of the time compared to typical python-based methods especially G-VAL, which also outperforms most of the others. While DIG and L-LC require the most time for detection since DIG implements a long timeout and retransmission mechanisms and L-LC requires 24 complete HTTPS connections to a single target. 
Second, when conducting a horizontal comparison of E-DoH across multiple datasets, variations in detection rates become evident with an increase in the number of probing targets. Specifically, on Dataset I with a detection count of only 131, E-DoH requires twice the time of the NSE method to complete the detection. However, on Dataset III with a detection count of 668, E-DoH takes approximately 62\% of the time compared to the NES method. This characteristic arises from the dynamic protocol negotiation, which serves to reduce the inefficiencies associated with redundant and invalid higher protocol handshakes. In scenarios with a larger number of potential targets and a lower proportion of DoH, the efficiency advantage of E-DoH becomes more evident, especially in wild detection.

\begin{table}
\centering
\footnotesize
\caption{Effective and traffic volume comparison}
\label{Tab_efficient_results}
\resizebox{\linewidth}{!}{
\begin{tblr}{
  cells = {c},
  cell{1}{2} = {c=3}{0.213\linewidth},
  cell{1}{5} = {c=3}{0.213\linewidth},
  cell{1}{8} = {c=3}{0.219\linewidth},
  vline{2-3,6} = {1}{},
  vline{2,5,8} = {1-11}{},
  hline{1,3,11} = {-}{},
}
Methods                                     & D I (131) &         &         & D II (144) &         &         & D III (668) &          &         \\
                                            & Success   & Time    & Bytes   & Success    & Time    & Bytes   & Success     & Time     & Bytes   \\
DIG                                 & 34.35\%   & 154m22s & 0.91MB  & 43.75\%    & 139m21s & 1.10MB  & 41.02\%     & 281m4s   & 4.41MB  \\
D-C53 & 34.35\%   & 9m2s    & 0.87MB  & 33.33\%    & 8m0s    & 1.08MB  & 26.35\%     & 98m42s   & 3.35MB  \\
G-NSE  & 46.56\%   & 1m27s   & 4.13MB  & 48.61\%    & 3m3s    & 5.53MB  & 42.66\%     & 14m21s   & 30.35MB \\
G-VAL   & 45.04\%   & 38m46s  & 3.70MB  & 47.22\%    & 38m22s  & 4.70MB  & 41.92\%     & 516m4s   & 19MB    \\
G-ALL  & 45.04\%   & 40m13s  & 7.83MB  & 47.22\%    & 41m25s  & 10.23MB & 41.92\%     & 530m25s  & 49.35MB \\
L-DE & 54.96\%   & 15m13s  & 2.79MB  & 52.78\%    & 14m48s  & 3.65MB  & 46.11\%     & 150m16s  & 12.12MB \\
L-LC              & 44.27\%   & 157m58s & 14.71MB & 51.39\%    & 170m35s & 20.44MB & 43.11\%     & 2158m19s & 77.01MB \\
E-DoH                                       & 45.80\%   & 2m39s   & 0.66MB  & 51.39\%    & 3m15s   & 0.75MB  & 43.26\%     & 8m53s    & 3.14MB  
\end{tblr}
}
\end{table}

\textbf{Recall Comparison} 
First, most methods exhibit roughly similar recall rate. Slight rate differences may stem from service instability and network conditions. Surprisingly, the L-LC method performs similar recall rates along with most others though it sends 24 packets to the same target.
Second, L-DE performs best in recall rates, since it introduces domain names as enhancement. Whereas DIG and D-C53 identify a lower number of services than other methods. DIG identifies fewer services since it only accepts TLS 1.3, while some methods only support a maximum of TLS 1.2. The low identification rate of the D-C53 method indicates the low statistical correlation between traditional DNS services and DoH services.

\textbf{Traffic Volume Comparison}
In theory, the dynamic protocol negotiation strategy only requires one single complete DoH connection. Although E-DoH and the L-LC method both employ path enumeration, the key difference lies in the fact that E-DoH's enumeration is built upon an established TLS connection, whereas L-LC requires the creation of a complete TLS connection for each path. Besides, other methods also establish a complete connection for each combination when enumerating TLS and HTTP versions. This results in a substantial amount of ineffective probing traffic. Therefore, other methods will consume 4-24 times more volume than E-DoH.
In practice, E-DoH accomplishes the detection task by transmitting an extremely small amount of volume. Overall, E-DoH only needs to transmit 4\%-20\% of the volume required by other methods to complete the detection task. Although the G-NSE demonstrates advantages in both efficiency and accuracy, it comes at the cost of significant volume consumption. This trend is also observed in the L-LC method. The experiment results further underscore the advantages of Dynamic Protocol Negotiation. The low traffic volume consumption is particularly crucial where cloud service providers currently adopt strict security strategies. 

\subsection{Thread Tuning}
\label{sec_thread}

\begin{figure}
\centerline{\includegraphics[width =0.6\linewidth]{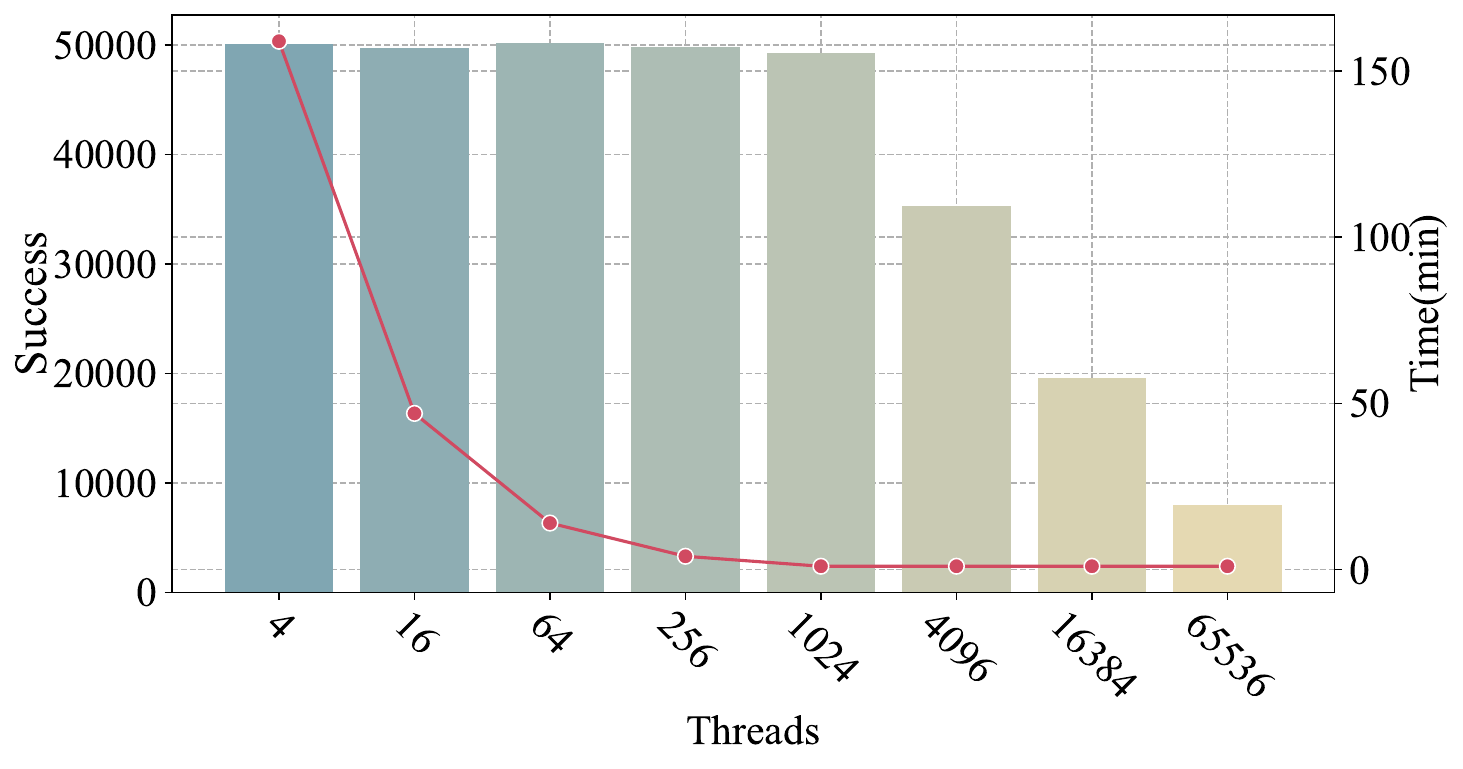}}
\caption{Tune results of E-DoH}
\label{Fig_BL_Tune}
\end{figure}

The number of threads simultaneously running has a significant impact on the efficiency of the E-DoH method. With a higher number of threads, there will be a greater likelihood of increased packet loss and timeouts due to CPU scheduling delays. To determine the most suitable number of threads, we conducted the threads tunning experiments in this section.
The settings of this experiment are as follows. First, the aforementioned three datasets are repeatedly sampled to create a collection of 10000 hosts, which provide sufficient and reliable DoH hosts for verification.
Second, the thread count increases exponentially in steps of 4, thus making it easier to identify performance inflection points with the growth of thread count.

The experimental results are depicted in \autoref{Fig_BL_Tune}. The bar chart represents the number of successful requests, and the line chart represents the time required for requests. 
First, regarding the success rate of detection,  E-DoH performs stable with no significant fluctuations when the number of threads is less than 1024. However, when the thread count exceeds 4096, there is a noticeable decrease in the success rate.
Second, regarding the time performance, the time required for E-DoH to complete the detection decreases rapidly as the thread count multiples increase when the number of threads is less than 1024. When the thread count exceeds 1024, the time stabilizes at around 1 minute. This phenomenon is likely due to the increased time needed for thread scheduling.
Therefore, based on the accuracy and time considerations of the above experimental integrated detection, we selected the number of threads to be 1024 for the subsequent experiments.

\section{Wild Exploration}
In this experiment, we aim to discover all publicly available DoH services in IPv4 space. The wild service detection process can be divided into three steps. The first step involves scanning the public addresses across the entire network on the port 443. Second, E-DoH is employed to probe hosts with open 443 ports for DoH service detection. Third, service status analysis is conducted based on collected responses and NS server logs. To carry out these experiments, we rented eight cloud hosts as vantage points, each with a primary configuration of 2 CPU cores, 4 GB of RAM, and a 5 Mbps bandwidth. Besides, the thread number is settled as 1024. They are located in Europe, Asia and North America.
This section shows the discovery results of the DoH and delves into an in-depth analysis of their service status. 

\subsection{Overview}
Due to privacy considerations and the potential impact on network services, not all DoH services respond to queries. Specifically, we categorize services as potential DoH services with the following two conditions: (1) the responses above HTTP can be successfully parsed as DNS message; (2) the requested domain name is recorded in the NS Server records.

\textbf{Detected Hosts}
Overview results are shown in \autoref{Tab_wild_main}. In the port probing procedure, around 50M hosts were found for opening port 443. Among these hosts, we identified 45930 as potential DoH hosts, 5007 of which are successfully responded. Compared to the previous study \cite{luoMeasurementEncryptedOpen2022,garciaLargeScaleAnalysis2022}, we identified more hosts, but the number of available DoH servers has not changed significantly. Consistent with our speculation, although some servers with limitations did not return valid results, they still sent requests to our backend and cached the results, this part of DoH services accounting for around 9.73\%.

\textbf{Protocol Negotiation}
The protocols used in negotiation of all detected hosts are shown in \autoref{Tab_wild_protocol}. Although \cite{hoffmanDNSQueriesHTTPS2018} suggested using TLS 1.3 and H2 for DoH deployment, only nearly half of the detected hosts follow this suggestion. In the TLS layer, there is no host default to using protocols below TLS 1.2 though we provide the option. Therefore, vulnerabilities associated with lower versions of the protocol can be effectively mitigated. In the HTTP layer, the vast majority of hosts (roughly 90\%) prefer using HTTP2 to provide services, while only a small number of DoH hosts explicitly choose HTTP1.1. 
Following our reasoning on deployment methods in section 5.1, we hypothesize that this part of services is deployed behind HTTP proxies. Since current mainstream DNS servers can directly support HTTP2 messages. Moreover, by forwarding HTTP information above the TLS layer directly to the corresponding backend services, it becomes more convenient to achieve compatibility with a diverse range of services in the backend.

\begin{table}[!htbp]
\centering
\caption{Wild DoH service identification results}
\label{Tab_wild_main}
\begin{tblr}{
  cells = {c},
  vline{3} = {-}{},
  hline{1,9} = {-}{0.08em},
  hline{2,8} = {-}{},
}
Status (\%) &          & Identified &       \\
TCP ERROR   & 57.83\%  & Success    & 5116  \\
TLS ERROR   & 17.05\%  & Empty      & 36452 \\
TIMEOUT     & 1.54\%   & Logged     & 4471  \\
NO PATH     & 23.50\%  &            &       \\
EMPTY       & 0.07\%   &            &       \\
SUCCESS     & 0.01\%   &            &       
\end{tblr}
\end{table}

\begin{table}[!htbp]
\centering
\caption{Dynamic protocol detail}
\label{Tab_wild_protocol}
\begin{tblr}{
  cells = {c},
  vline{4} = {-}{},
  vline{4} = {2}{-}{},
  hline{1,7} = {-}{0.08em},
  hline{2} = {-}{},
}
Protocol &    &         & Path      &         \\
TLS 1.2  & H2 & 32.50\% & dns-query & 76.99\% \\
TLS 1.2  & H1 & 4.74\%  & /         & 20.94\% \\
TLS 1.3  & H2 & 47.33\% & NO PATH   & 1.32\%  \\
TLS 1.3  & H1 & 15.42\% & resolve   & 0.42\%  \\
         &    &         & doh       & 0.33\%  
\end{tblr}
\end{table}

\subsection{Response Analysis}
Resolver-side hijacking is a typical method of DNS hijacking. Attackers use this method to replace correct resolution records with other records in order to direct users to undesired websites. This is achieved for various purposes, including gaining economic benefits, conducting attacks, or implementing access control.
For specific, correctness analysis can be achieved by checking the consistency between the returned result record and the zone file record. Due to the wildcard records settled in the zonefile, we can assess the correctness of the requested records conveniently. The wildcard records can map different domain records to the same IP address. In our zonefile, we configured a wildcard record \emph{*.001.uniquetest.today} pointing to the record 255.255.255.251, which serves as the basis for the correct record. Therefore, we only need to check if each returned record is equal to 255.255.255.251, without setting a separate correct record for each unique query. 

According to our analysis, nearly 98\% of the DoH hosts with valid responses returned the correct answer. When compared to traditional DNS services, there is less hijacking behavior observed on DoH services. This highlights the higher service quality of our discovered DoH.
Besides, we manually analyzed the incorrect answer with open threat intelligence and the web page. In total, these incorrect answers are categorized into four types, malicious, advertise, blocked and honeypot.
In the first category, some of the answers try to redirect users to multiple web pages with a large amount of Javascript. We suspect that this hijacking method is motivated by the pursuit of economic gains, which is similar to traditional DNS hijacking. Another part of these answers tries to redirect users to C\&C servers indicated by the hybrid-analysis platform. We suspect that these DoH hosts are involved in the control chain of the attackers. 
In the second category, all of the answers try to redirect users to secure vendors' homepages for advertising purposes. In the third category, the answers are returned as 0.0.0.0 or redirect users to an empty web page. In the fourth category, the answer contains CNAME records indicating it is a sinkhole of a university.

\subsection{Parsing Dependency}
DNS resolvers can be classified into two types based on the query approach, the forwarding resolver and the recursive resolver. Forwarding resolvers perform recursive queries to other resolvers, and recursive resolvers perform iterative queries. Based on the backend logs recorded by our controlled NS server, the discovered DoH resolver can also be classified into these two types. 

\textbf{Forwarding vs Recursive}
In backend logging, each log primarily consists of two components, the recorded domain and the host with the $IP_{record}$ that made the request. Since the probing target is embedded in the requested domain as $IP_{request}$, the request IP can be extracted from the recorded domain. Hence, by checking the consistency of $IP_{request}$ and $IP_{record}$, the type of the probing target $IP_{request}$ can be determined. More specifically, if $IP_{request}$ and $IP_{record}$ are the same, the request host $IP_{request}$ is a recursive DoH resolver. Otherwise, the request host $IP_{request}$ is a forwarding DoH resolver.
According to our analysis, only 8\% of the logged resolvers are recursive resolvers, and nearly 92\% are forwarding resolvers. Besides, we also found that there are 8 inter resolvers, which means that they both perform iterative queries and recursive queries at the same time. 

\textbf{Recursive Resolver}
Since most of the resolvers are recursive resolvers, we try to dig out more about their service status.
First, we find that the ownership of these recursive resolvers is quite dispersed. Since these servers typically only have the port 53 open, we cannot identify the service provider as we did with DoH services through certificates. Therefore, we conduct an analysis based on the organization associated with IP addresses using the MaxMind database. The top three organizations providing resolving IPs are Google, CLOUDFLARE, and OPENDNS, as shown in Appendix \autoref{Fig_Hidden_pie}.
Besides, we perform additional traditional DNS requests to these services since the iterative queries are mainly based on 53 ports. We find that most recursive resolvers do not provide DNS resolution services to the external world. After sending DNS requests to the port 53 of these recursive resolvers, the results can be seen in Appendix \autoref{Tab_hidden_respons}. It can be observed that the majority of indirect resolvers return TIMEOUT as the result. We speculate that the existence of these resolvers primarily aims at building multi-level caching services rather than directly providing resolution services.

\begin{figure*}[!htbp]
\centering
\subfigure[ Cumulative cure of recursive resolver]{
\label{Fig_cum_recur}
\includegraphics[width=0.45\linewidth]{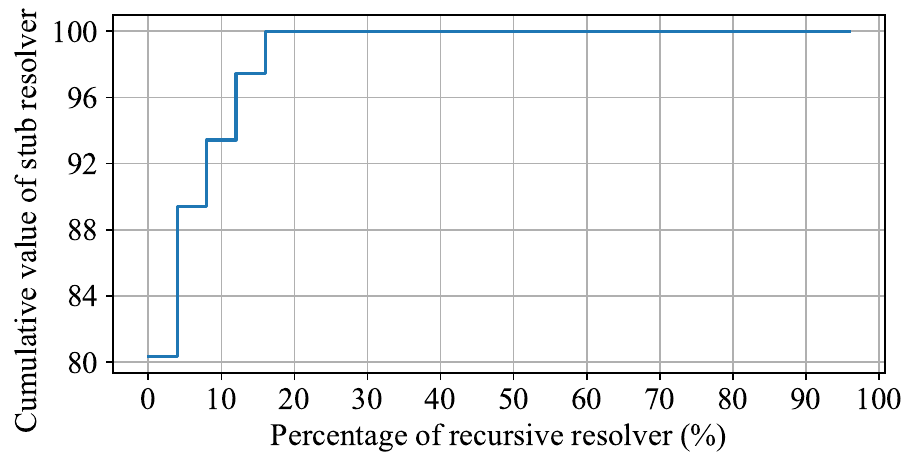}
}
\subfigure[ Cumulative cure of forwarding resolver]{
\label{Fig_cum_stub}
\includegraphics[width=0.45\linewidth]{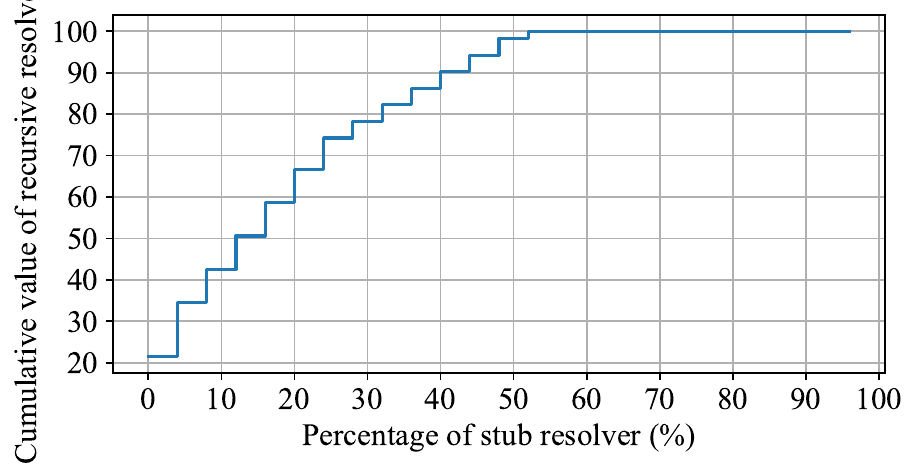}
}
\caption{Cumulative curves} 
\label{Fig_cum}
\end{figure*}

\textbf{Cumulative Analysis}
To reveal the quantity relationship between the forwarding resolvers and the recursive resolvers, we conducted a cumulative analysis. Results are shown in \autoref{Fig_cum}. 
It can be found from \autoref{Fig_cum_recur} that around 90\% of the forwarding DoH resolvers rely on only 5\% of the recursive resolver. Based on the organization analysis of these IPs with the Maxmind database, these recursive resolvers mostly belong to Google. 
This shows that within the DoH ecosystem, machines performing actual iterative queries remain concentrated in the resolvers of a few large organizations. While this highly centralized resolution structure can improve the efficiency of a single iterative query result, it poses a significant impact on forwarding resolvers in the event of a recursive resolver being hijacked or polluted.
Second, 
it can be shown from \autoref{Fig_cum_stub} that 50\% of the recursive resolvers are requested by 15\% of the forwarding DoH servers. This means that some forwarding DoH forwarders rely on multiple recursive resolvers. We speculate that by querying multiple recursive servers, the resolvers can compare and verify the resolution results to ensure that the provided IP addresses are accurate and consistent. 
Besides, the conflict caused by different update times of the domain can also be reduced. Thus the reliability and stability of resolution can be improved. 

\textbf{Dependency Structure}
\begin{figure}[!htbp]
\centerline{\includegraphics[width =0.45\linewidth]{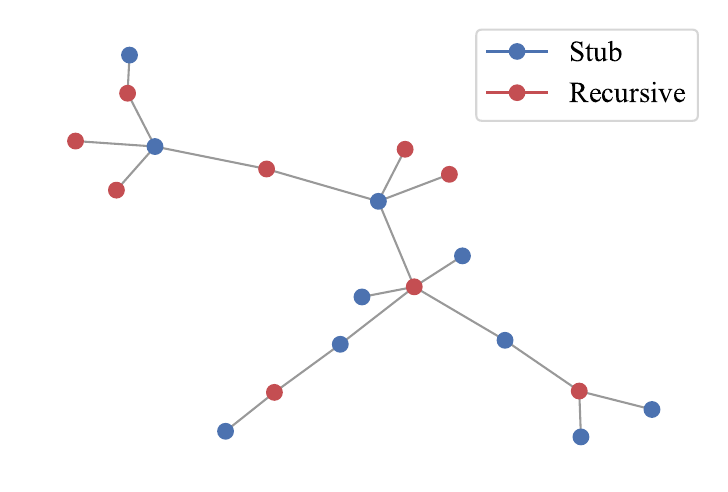}}
\caption{Sample partial service resolution dependency connectivity component}
\label{Fig_Depend}
\end{figure}
From the cumulative analysis, it can be acquired that the dependency of the direct resolvers and hidden resolvers are complex. To reveal more detail of the dependency, we correlate all detected DoH resolvers to graphs. 
As a primary result, DoH resolvers can be composed into several independent connected components. Each component reflects the tight dependency of involved DoH resolvers. An example of a small-scale connectivity subgraph is shown in \autoref{Fig_Depend}. 
The significance of the dependency graph is to analyze risk propagation and make corresponding countermeasures. For example, the TsuKing \cite{10.1145/3576915.3616668} attack that leverages the dependency of the DNS structure for DNS amplification attack can be analyzed with the dependency structure. Furthermore,
if there is DNS pollution in one of the nodes, such erroneous DNS caches can be passed on to other resolvers through the resolution dependency resulting in a large impact. 

\section{Discussion}
\subsection{Ethical Consideration}
There are usually potential ethical concerns associated with high rate network probing. This paper identifies two main potential ethical risks. At the network level, both port scanning and service probing processes can potentially affect the source network and the network of the target ISP. To address this issue, we have taken several measures. First, we randomly order the probing targets to evenly distribute the probing load over time, reducing the likelihood of initiating a large number of requests to a single target in a short period. Second, we distribute scanning tasks across multiple leased hosts to minimize the pressure on the source network. Third, we employ a low scanning rate to avoid exhausting bandwidth on the network paths.
At the host level, service probing can also exert pressure on both the probing host and the target host. First, we rent several probing hosts and pay for them. This allows us to legitimately utilize their resources. Second, we only initiate one TCP session for each target host with an open 443 port. This ensures a limited impact on the probing host for their external services. Additionally, the filtering strategy of E-DoH enables the early termination of sessions for most probing targets after the TLS handshake. 
Therefore, in summary, the impact of our approach on existing services is negligible.

\subsection{Shortage of Our Method}

Since DoH services rely on the HTTP protocol, they need to first go through the authentication process of HTTP. HTTP utilizes domain names and paths to locate requested resources. This mechanism can also serve as an authentication mechanism when deployed accordingly. Therefore, for DoH services that require specific subdomains and specific paths, although E-DoH can detect them based on specific configurations, it is unable for E-DoH to proactively discover these specific settings. In the case of such services, we classify them as non-public DoH services, which can be discovered through a combination of passive observation, domain discrepancy analysis, and path enumeration methods.

\section{Conclusion}
The detection of DoH services is currently a hot topic in the field of network measurement. This paper achieves the efficient detection of publicly available DoH services primarily through optimized detection mechanisms and probing implementation. Additionally, this paper provides a multi-faceted analysis of current public DoH services. From an overarching perspective, current DoH services exhibit complex dependencies while maintaining a high level of service quality.

\section*{Acknowledgment}

We thank the anonymous reviewers for their constructive suggestions, comments, and valuable insights, which helped us improve the paper. 
\nocite{*}
\bibliographystyle{splncs04}
\bibliography{Thesis}

\begin{subappendices}
\renewcommand{\thesection}{\Alph{section}}%

\newpage
\section{Appendix}
\begin{figure}[!htbp]
\centerline{\includegraphics[width =0.6\linewidth]{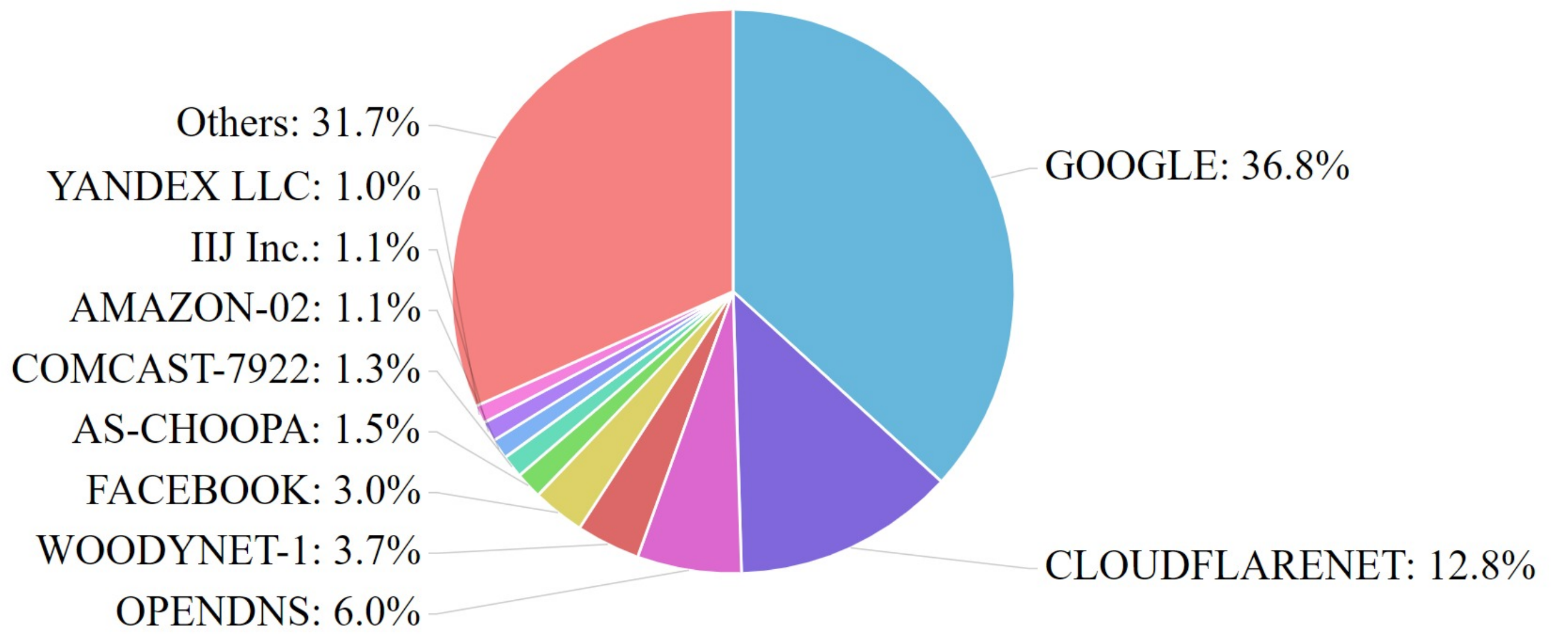}}
\caption{Providers of the hidden resolvers}
\label{Fig_Hidden_pie}
\end{figure}

\begin{table}
\centering
\caption{Response statistics of the hidden resolvers}
\label{Tab_hidden_respons}
\begin{tabular}{ccc} 
\hline
Response & Percentage     \\ 
\hline\hline
TIMEOUT     & 93.7\%  \\ 
NOERROR     & 4.0\%   \\ 
REFUSED     & 2.3\%   \\
\hline
\end{tabular}
\end{table}

\end{subappendices}
\end{document}